\documentclass[12pt,a4paper,confbib]{conference}

\usepackage{fancyhdr}
\usepackage{graphicx,amsmath,amssymb,cite}
\usepackage{multind}
\makeindex{author} \makeindex{subject}

\newcommand{\beq}{\begin{equation}}
\newcommand{\eeq}[1]{\label{#1}\end{equation}}
\newcommand{\eeqn}{\end{equation}}

\newcommand{\beqa}{\begin{eqnarray}}
\newcommand{\eeqa}[1]{\label{#1}\end{eqnarray}}
\newcommand{\eeqan}{\end{eqnarray}}

\let\bar=\overbar

\newcommand{\Dslash}{\not{\hbox{\kern-4pt $D$}}}
\newcommand{\dslash}{\not{\hbox{\kern-2pt $\del$}}}

\newcommand{\msb}{{\bar{\ssstyle M \kern -1pt S}}}

\begin{document}

\Chapter{Dynamics, Symmetries \\ and Hadron Properties}
           {Dynamics, Symmetries \& Hadrons}{I.C.~Clo\"et \textit{et al}.}
\vspace{-5 cm}\includegraphics[width=6 cm]{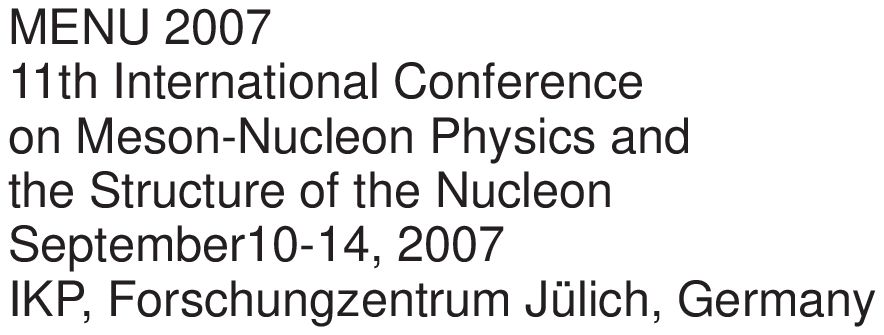}
\vspace{3 cm}

\addcontentsline{toc}{chapter}{{\it I.C.~Clo\"et \textit{et al}.}} \label{authorStart}

\begin{raggedright}

{\it I.C.~Clo\"et,$^\ast$ A.~Krassnigg\,$^\dagger$ and C.D. Roberts\,$^\ast$}
\index{author}{Clo\"et, I.C., \textit{et al}.}
\medskip

$^\ast$\,\parbox[t]{0.9\textwidth}{Physics Division, 
Argonne National Laboratory,\\
Argonne, Illinois 60439,
United States of America}
\smallskip

$^\dagger$\,\parbox[t]{0.9\textwidth}{Institut f\"ur Physik,
Karl-Franzens-Universit\"at Graz, \\
A-8010 Graz, 
Austria}

\bigskip\bigskip

\end{raggedright}

\begin{center}
\textbf{Abstract}
\end{center}
We provide a snapshot of Dyson-Schwinger equation applications to the theory and phenomenology of hadrons.  Exact results for pseudoscalar mesons are highlighted, with details relating to the $U_A(1)$ problem.  Calculated masses of the lightest $J=0,1$ states are discussed.  We recapitulate upon studies of nucleon properties and give a perspective on the contribution of quark orbital angular momentum to the spin of a nucleon at rest.

\section{Introduction}
Numerous salient features exhibited by the physics of mesons and nucleons arise nonperturbatively in QCD.  Two phenomena strike one immediately: confinement and dynamical chiral symmetry breaking (DCSB).  DCSB is the better understood of these emergent phenomena; e.g., it explains the origin of constituent-quark masses and underlies the success of chiral effective field theory.  Confinement, on the other hand, remains only an empirical fact; viz., colored objects have not hitherto been observed in isolation.  A fact too often ignored is that the potential between infinitely heavy quarks measured in numerical simulations of quenched lattice-QCD -- the static potential -- is not related in any known way to light-quark confinement.

\section{DCSB}
Understanding DCSB within QCD proceeds from the gap equation \cite{Maris:1997hd}; namely, the Dyson-Schwinger equation (DSE) for the dressed-quark propagator:
\begin{equation}
S(p)^{-1} =  Z_2 \,(i\gamma\cdot p + m^{\rm bm}) + Z_1 \int^\Lambda_q\! g^2 D_{\mu\nu}(p-q) \frac{\lambda^a}{2}\gamma_\mu S(q) \Gamma^a_\nu(q,p) , \label{gendse}
\end{equation}
where $\int^\Lambda_q$ represents a Poincar\'e invariant regularization of the integral, with $\Lambda$ the regularization mass-scale, $D_{\mu\nu}$ is the dressed-gluon propagator, $\Gamma_\nu$ is the dressed-quark-gluon vertex, and $m^{\rm bm}$ is the quark's $\Lambda$-dependent bare current-mass.  The vertex and quark wave-function renormalization constants, $Z_{1,2}(\zeta^2,\Lambda^2)$, depend on the gauge parameter.  

The solution of Eq.\,(\ref{gendse}) can be written:
\begin{eqnarray} 
 S(p) & =&  \frac{1}{i \gamma\cdot p \, A(p^2,\zeta^2) + B(p^2,\zeta^2)} =
\frac{Z(p^2,\zeta^2)}{i\gamma\cdot p + M(p^2)}\,,
%
\label{Sgeneral}
\end{eqnarray} 
wherein $Z(p^2;\zeta^2)$ is the wave-function renormalization and $M(p^2)$ is the mass function.  The latter is independent of the renormalization point, $\zeta$.  The dressed propagator is obtained from Eq.\,(\ref{gendse}) augmented by the renormalization condition
$\left.S(p)^{-1}\right|_{p^2=\zeta^2} = i\gamma\cdot p +
m(\zeta^2)\,,$
where $m(\zeta^2)$ is the running mass: 
$Z_2(\zeta^2,\Lambda^2) \, m^{\rm bm}(\Lambda) = Z_4(\zeta^2,\Lambda^2) \, m(\zeta^2)\,,$
with $Z_4$ the Lagrangian-mass renormalization constant.  In QCD the chiral limit is strictly defined by \cite{Maris:1997hd}:
$Z_2(\zeta^2,\Lambda^2) \, m^{\rm bm}(\Lambda) \equiv 0 \,, \forall \Lambda^2 \gg \zeta^2 ,$
which states that the renormalization-point-invariant cur\-rent-quark mass $\hat m = 0$. 

Only in the chiral limit is it possible to unambiguously define the gauge invariant vacuum quark condensate in terms of $S(p)$ \cite{Maris:1997hd,Langfeld:2003ye,Chang:2006bm}.  This emphasizes that gauge covariant quantities contain gauge invariant information.  The condensate is the order parameter most commonly cited in connection with DCSB.  Nonetheless, $M(p^2)$ is a more fundamental indicator: the condensate is only a small part of the information it contains.

In perturbation theory it is impossible in the chiral limit to obtain $M(p^2)\neq 0$: the generation of mass from \emph{nothing} is an essentially nonperturbative phenomenon.  On the other hand, it is a longstanding prediction of nonperturbative DSE studies that DCSB will occur so long as the integrated infrared strength possessed by the gap equation's kernel exceeds some critical value \cite{Roberts:1994dr}.  There are strong indications that this condition is satisfied in QCD \cite{Bowman:2005vx,Bhagwat:2006tu}.  It follows that the quark-parton of QCD acquires a momentum-dependent mass function, which at infrared momenta is $\sim 100$-times larger than the current-quark mass.  This effect owes primarily to a dense cloud of gluons that clothes a low-momentum quark \cite{Bhagwat:2007vx}.

A great deal has been learnt from the gap equation alone.  To highlight only one recent example \cite{Chang:2006bm}, realistic \textit{Ans\"atze} for the the gap equation's kernel indicate that there is a critical current-quark mass, $\hat m_{\rm cr}$, above which $M(p^2)$ does not possess an expansion in $\hat m$ around its chiral-limit value.  For a pion-like meson constituted from a quark, $f$, with mass $\hat m_{\rm cr}$ and an equal-mass different-flavor antiquark, $\bar g$, $m_{\bar g f}^{0^-} = 0.45\,$GeV.  Since physical observables, such as the leptonic decay constant, are expressed in terms of $M(p^2)$, it follows that a chiral expansion is meaningful only for $(m_{\bar g f}^{0^-})^2 \lesssim 0.2\,$GeV$^2$.  This entails, e.g., that it is only valid to employ chiral perturbation theory to fit and extrapolate results from lattice-regularized QCD when the simulation parameters provide for $m_\pi^2 \lesssim 0.2\,$GeV$^2$.  Lattice results at larger pion masses are not within the domain of convergence of chiral perturbation theory. 

\section{Mesons}
For $0^-$ mesons the axial-vector Ward-Takahashi identity is of fundamental importance, and recently its implications for neutral pseudoscalars and $\eta$-$\eta^\prime$ mixing have been elucidated \cite{Bhagwat:2007ha}.  In the general case the identity is written 
\begin{equation}
P_\mu \Gamma_{5\mu}^a(k;P) ={\cal S}^{-1}(k_+) i \gamma_5 {\cal F}^a 
+ i \gamma_5 {\cal F}^a {\cal S}^{-1}(k_-)
- 2 i {\cal M}^{ab}\Gamma_5^b(k;P)  - {\cal A}^a(k;P)\,,
\label{avwti}
\end{equation}
wherein: $\{{\cal F}^a | \, a=0,\ldots,N_f^2-1\}$ are the generators of $U(N_f)$; the dressed-quark propagator ${\cal S}=\,$diag$[S_u,S_d,S_s,S_c,S_b,\ldots]$; ${\cal M}(\zeta)$ is the matrix of running current-quark masses and
${\cal M}^{ab} = {\rm tr}_F \left[ \{ {\cal F}^a , {\cal M} \} {\cal F}^b \right],$
where the trace is over flavour indices.  The inhomogeneous axial-vector vertex in Eq.\,(\ref{avwti}), $\Gamma_{5\mu}^a(k;P)$, where $P$ is the total momentum of the quark-antiquark pair and $k$ the relative momentum, satisfies a Bethe-Salpeter equation (BSE), and likewise the pseudoscalar vertex, $\Gamma_5^b(k;P)$.  The final term in Eq.\,(\ref{avwti}) expresses the axial anomaly.  It involves 
\begin{equation}
{\cal A}_U(k;P) = \!\!  \int\!\! d^4xd^4y\, e^{i(k_+\cdot x - k_- \cdot y)} N_f \left\langle  {\cal F}^0\,q(x)  \, {\cal Q}(0) \,   \bar q(y) 
\right\rangle, \label{AU}
\end{equation}
wherein the matrix element represents an operator expectation value in full QCD and
${\cal Q}(x) = i \frac{\alpha_s }{8 \pi} \, \epsilon_{\mu\nu\rho\sigma} F^a_{\mu\nu} F^a_{\rho\sigma}(x) 
= \partial_\mu K_\mu(x)$
is the topological charge density operator, where $F_{\mu\nu}^a$ is the gluon field strength tensor.\footnote{NB.\ While ${\cal Q}(x)$ is gauge invariant, the associated Chern-Simons current, $K_\mu$, is not.  Thus in QCD no physical state can couple to $K_\mu$.  Hence, physical states cannot provide a resolution of the so-called $U_A(1)$-problem; namely, they cannot play any role in ensuring that the $\eta^\prime$ is not a Goldstone mode.}

In considering the $U_A(1)$-problem one need only focus on the case ${\cal A}^{0} \neq 0$ because if that is false, then following Ref.\,\cite{Maris:1997hd} it is clear that the $\eta^\prime$ is certainly a Goldstone mode.  ${\cal A}^0$ is a pseudoscalar vertex and can therefore be expressed
\begin{eqnarray}
\nonumber 
{\cal A}^0(k;P) & = & {\cal F}^0\gamma_5 \left[ i {\cal E}_{\cal A}(k;P) + \gamma\cdot P {\cal F}_{\cal A}(k;P)  \right. \\
&& \left. +\gamma\cdot k k\cdot P {\cal G}_{\cal A}(k;P) + \sigma_{\mu\nu} k_\mu P_\nu {\cal H}_{\cal A}(k;P)\right].
\end{eqnarray}
Equation\,(\ref{avwti}) can now be used to derive a collection of chiral-limit, pointwise Goldberger-Treiman relations, important amongst which is the identity
$2 f_{\eta^\prime} E_{\eta^\prime}(k;0) = 2 B_{0}(k^2) - {\cal E}_{\cal A}(k;0)\,,$
where $B_0(k^2)$ is obtained in solving the chiral-limit gap equation.  It is plain that if 
\begin{equation}
\label{calEB}
{\cal E}_{\cal A}(k;0) = 2 B_{0}(k^2) \,,
\end{equation}
then $f_{\eta^\prime} E_{\eta^\prime}(k;0) \equiv 0$.  This being true, then the homogeneous Bethe-Salpeter equation for the $\eta^\prime$ does not possess a massless solution in the chiral limit.  The converse is also true.  Hence Eq.\,(\ref{calEB}) is a necessary and sufficient condition for the absence of a massless $\eta^\prime$ bound-state.  The chiral limit is being discussed, in which case $B_{0}(k^2) \neq 0$ if, and only if, chiral symmetry is dynamically broken.   Thus the absence of a massless $\eta^\prime$ bound-state is only assured through the existence of an intimate connection between DCSB and an expectation value of the topological charge density.  A  relationship between the mechanism underlying DCSB and the absence of a ninth Goldstone boson was also discussed in Ref.\,\cite{Dorokhov:2003kf}.

Reference\,\cite{Bhagwat:2007ha} also derives corollaries, amongst which are mass formulae for neutral pseudoscalars, and presents an \textit{Ansatz} for the Bethe-Salpeter kernel that enables their illustration.  The model is elucidative and phenomenologically efficacious; e.g., it predicts $\eta$--$\eta^\prime$ mixing angles of $\sim - 15^\circ$ and $\pi^0$--$\eta$ angles of $\sim 1^\circ$; and suggests a strong neutron-proton mass difference of $0.75\,(m_d - m_u)$.

The use of DSEs to study meson phenomena is empowered by the existence of a systematic, nonperturbative and symmetry-preserving truncation scheme \cite{Munczek:1994zz,Bender:1996bb}.  It means that exact results, such as those indicated above, and others related to radial excitations and/or hybrids \cite{Holl:2004fr,Holl:2005vu,McNeile:2006qy}, and heavy-light \cite{Ivanov:1998ms} and heavy-heavy mesons \cite{Bhagwat:2006xi}, can be proved and illustrated.  

\begin{table}[t]
\caption{\label{masses} Masses (GeV) of the lightest $J=0,1$ states produced by the rainbow-ladder DSE truncation of Refs.\,\protect\cite{Maris:1997tm,Maris:1999nt} with the parameter values: $\omega = 0.4\,$GeV, $\omega D = (0.72\,$GeV$)^3$; and current-quark masses $m_{u,d}(1\,{\rm GeV}) = 5.45\,$MeV, $m_{s}(1\,{\rm GeV}) = 125\,$MeV.  The rainbow-ladder kernel gives ideal flavour mixing for all states.
See the text for further discussion.}
\begin{center}
\begin{tabular}{l||l|l||l||l|l||l|l|l} \hline \hline
\rule{0ex}{3ex} $J^{PC}$ & $0^{-+}$ & $1^{--}$ & $0^{++}$ & $1^{+-}$ & $1^{++}$ & $0^{--}$ & $1^{-+}$ & $0^{+-}$ \\
$\bar u u$ & 139 & ~740 & ~670 & ~830  & ~900 & ~860 & 1000 & 1040 \\
$\bar s s$ & 695 & 1065 & 1080 & 1165 & 1240 & 1170 & 1310 & 1385
\\
\hline \hline
\end{tabular} 
\end{center}
\end{table}

In the latter connection, the renormalization-group-improved rainbow-ladder truncation of the gap and Bethe-Salpeter equations introduced in Refs.\,\cite{Maris:1997tm,Maris:1999nt} has been widely employed.  To exemplify that, in Table\,\ref{masses} we report calculated results for the masses of the lightest $J=0,1$ states \cite{pmprivate}.  It is true in general that the truncation is accurate for the $0^{-+}$ and $1^{--}$ light-quark meson ground states.  In these channels it can be seen algebraically that contributions beyond rainbow-ladder largely cancel between themselves owing to Eq.\,(\ref{avwti}) \cite{Bender:1996bb,Bender:2002as,Bhagwat:2004hn,Matevosyan:2006bk}.  The remaining columns in the table deserve special attention because they show clearly the path toward improvement.

Terms beyond the rainbow-ladder truncation are known to add constructively in the $0^{++}$ channel \cite{Roberts:1996jx}.  Hence the leading order truncation is \emph{a priori} not expected to provide a good approximation.  Further understanding is provided by an exploration of the contribution from two-pion intermediate states to the mass and width of this lowest-mass scalar.  A rudimentary analysis shows that a realistic description is attainable therewith \cite{Holl:2005st}; viz., it gives a pole position $\surd s_\sigma = (0.578 - i \, 0.311)\,$GeV.  This is not the end of the scalar story but it is a sensible path to follow, in particular because a QCD-level mechanism is precisely specified.\footnote{In rainbow-ladder truncation, at least up to the c-quark mass, the ordering of meson masses is $0^{-+}(1S) < 0^{++}(1S) < 0^{-+}(2S) < 0^{++}(2S)$ \protect\cite{Krassnigg:2006ps}.}

Compared with experiment, the masses of the axial-vector mesons $1^{+\pm}$ are poorly described by the rainbow-ladder truncation: $\sim 400\,$MeV of repulsion is missing from the kernel.  A cruder model does better \cite{Burden:1996nh,Bloch:1999vka}.  The latter studies and a more recent analysis \cite{Watson:2004kd} indicate that at least part of the defect owes to the absence of spin-flip contributions at leading-order.  Such contributions appear at all higher orders and are enhanced by the strongly dressed quark mass function.  It is in this way that the meson spectrum can be used to probe the long-range part of the light-quark interaction and thereby to chart the nonperturbative behavior of QCD's $\beta$-function. 

The last three columns describe systems with so-called exotic quantum numbers.  Of course, these states are exotic only in the context of the naive constituent quark model.  In QCD they correspond simply to interpolating fields with some gluon content and are easily accessible via the BSE \cite{Burden:2002ps}.  Nonetheless, while the rainbow-ladder truncation binds in these channels, the shortcomings encountered in the $1^{+}$ channels are also evident here, for much the same reasons.  Reliable predictions for the masses of such states will only be obtained once improved kernels are developed.  At the very least, one must have reliable predictions for axial-vector masses before drawing any conclusions about the so-called exotics.

One might pose the question of whether, in the context of bound-state studies in which model assumptions are made regarding the nature of the long-range interaction between light quarks, anything is gained by working solely with Schwinger functions.  This means, in part, constraining oneself to work only with information obtained from the DSEs at spacelike momenta.\footnote{Lattice-regularized QCD provides a background to this question.  That approach is grounded on the Euclidean space functional integral.  Schwinger functions; i.e., propagators and vertices at spacelike momenta, are all that it can directly provide.  It can only be useful if methods are found so that the question can be answered in the affirmative.}  According to a recent study \cite{Bhagwat:2007rj} the answer is no.  It analysed the capacity of Schwinger functions to yield information about bound states, and established that for the ground state in a given channel the mass and residue are accessible via rudimentary methods.  However, simple methods cannot provide dependable information about more massive states in a given channel.  Indeed, there is no easy way to extract such information.  An approach based on a correlator matrix can be successful but only if the operators are carefully constructed so as to have large overlap with states of interest in a given channel and statistical errors can be made small; viz., $\sim 1$\%.  While it is possible in principle to satisfy these constraints, doing so is labor intensive and time consuming.  That is only justified in the absence of model-dependence.

\section{Nucleons}
The discussion of DSEs at MENU04 did not describe the study of nucleons, stating only that it was feasible \cite{Holl:2004un}.  Material progress has been made in the interim.  We now possess a level of expertise roughly equivalent to that we had with mesons approximately ten years ago; viz., phenomenology constrained by the significant body of knowledge we have gained in meson applications.  

The nucleon appears as a pole in a six-point quark Green function.  The pole's residue is proportional to the nucleon's Faddeev amplitude, which is obtained from a Poincar\'e covariant Faddeev equation that adds-up all possible quantum field theoretical exchanges and interactions that can take place between three dressed-quarks.  Poincar\'e covariance is crucial because modern experimental facilities employ large momentum transfer reactions.

A tractable truncation of the Faddeev equation is based \cite{Cahill:1988dx} on the observation that an interaction which describes mesons also generates colour-$\bar 3$ diquark correlations \cite{Cahill:1987qr}.  For ground state octet and decuplet baryons the dominant correlations are $0^+$ and $1^+$ diquarks because, e.g.: the associated mass-scales are smaller than the baryons' masses \cite{Hanhart:1995tc,Burden:1996nh,Maris:2002yu}, namely (in GeV)
$m_{[ud]_{0^+}} = 0.7 - 0.8$,
$m_{(uu)_{1^+}}=m_{(ud)_{1^+}}=m_{(dd)_{1^+}}=0.9 - 1.0$;
and the electromagnetic size of these correlations is less than that of the proton \cite{Maris:2004bp} -- $r_{[ud]_{0^+}} \approx 0.7\,{\rm fm}$, which implies  $r_{(ud)_{1^+}} \sim 0.8\,{\rm fm}$ based on the $\rho$-meson/$\pi$-meson radius-ratio \cite{Maris:2000sk,Bhagwat:2006pu}.

The Faddeev equation's kernel is completed by specifying that the quarks are dressed, with two of the three dressed-quarks correlated always as a colour-$\bar 3$ diquark.  Binding is then effected by the iterated exchange of roles between the bystander and diquark-participant quarks.  A Ward-Takahashi-identity-pre\-ser\-ving electromagnetic current for the baryon thus constituted is subsequently derived~\cite{Oettel:1999gc}.  It depends on the electromagnetic properties of the axial-vector diquark correlation.

A study of the nucleon's mass and the effect on this of a pseudoscalar meson cloud are detailed in \cite{Hecht:2002ej}.  Lessons learnt were employed in a series of studies of nucleon properties, including form factors \cite{Alkofer:2004yf,Holl:2005zi,Bhagwat:2006py,Holl:2006zw}.  The calculated ratio $\mu_p G_E^p(Q^2)/G_M^p(Q^2)$ passes through zero at $Q^2\approx 6.5\,$GeV$^2$ \cite{Holl:2005zi}.  For the neutron, in the neighbourhood of $Q^2=0$, $\mu_n\, G_E^n(Q^2)/G_M^n(Q^2) = - \frac{r_n^2}{6}\, Q^2$, where $r_n$ is the neutron's electric radius \cite{Bhagwat:2006py}.  The evolution of $\mu_p G_E^p(Q^2)/G_M^p(Q^2)$ and $\mu_n G_E^n(Q^2)/G_M^n(Q^2)$ on $Q^2\gtrsim 2\,$GeV$^2$ are both primarily determined by the quark-core of the nucleon.  While the proton ratio decreases uniformly on this domain \cite{Alkofer:2004yf,Holl:2005zi}, the neutron ratio increases steadily until $Q^2\simeq 8\,$GeV$^2$ \cite{Bhagwat:2006py}.  A comparison of the Pauli/Dirac form factor ratios for the neutron and proton is presented in Ref.\,\cite{Bhagwat:2007vx}.

Of significant interest is the distribution of an hadron's \emph{spin} over the quark constituents and their angular momentum.  In a Poincar\'e covariant approach that can be calculated in any frame.  The rest frame is physically most natural.  The pion was considered in Ref.\,\cite{Bhagwat:2006xi}. The answer is more complicated for the spin-$\frac{1}{2}$ nucleon.  In the truncation just described  a nucleon's Faddeev wave-function is expressed through eight scalar functions: no more are needed, no number fewer is complete.  Two are associated with the $0^+$ diquark correlation: ${\cal S}_{1,2}$, and six with the $1^+$ correlation: ${\cal A}_{1,\ldots,6}$.  In the rest frame in this basis one can derive the following ``good'' angular momentum and spin assignments, which add vectorially to give a $J=\frac{1}{2}$ nucleon:\footnote{Equation~(3.35) or Ref.\,\protect\cite{Oettel:1998bk} contradicts Fig.\,6 of that reference.  Equation~(\protect\ref{LS}) herein describes the correct assignments.}
\begin{eqnarray}
&&
\begin{array}{c|c|c|c}
L=0\,, S=\frac{1}{2} & L=1\,,S=\frac{1}{2} &  L=1\,,S=\frac{3}{2} & L=2 \,, S=\frac{3}{2}\\\hline
{\cal S}_1\,,{\cal A}_{2}\,, {\cal B}_1 & {\cal S}_2\,, {\cal A}_1\,, {\cal B}_2\,, & {\cal C}_2& {\cal C}_1
\end{array}\;, \\
&&
\begin{array}{cclccl}
 B_1 & =& \frac{1}{3} A_3 + \frac{2}{3} A_5\,, &
 B_2 & =& \frac{1}{3} A_4 + \frac{2}{3} A_6\,, \\
 C_1 & =& A_3 - A_5\,, &
 C_2 & =& A_4 - A_6 \,.
\end{array} \label{LS}
\end{eqnarray}
These assignments are straightforward to demonstrate and understand; e.g, in the rest frame of a relativistic constituent quark model the ${\cal S}_{1,2}$ terms correspond, respectively, to the upper and lower components of the nucleon's spinor.

To exhibit the importance of the various $L$-$S$ correlations within the nucleon's Faddeev wave-function we report the breakdown of contributions to the nucleon's canonical normalization:\footnote{The entry in location ${\cal S}_1 \otimes {\cal S}_1$ indicates the integrated contribution associated with ${\cal S}_1^2$.  The entries are reweighted such that the sum of the squares of the entries equals one.  Positions without an entry are zero to two decimal places.}
\begin{equation}
\begin{array}{l|rrr|rrr|r|r}
& {\cal S}_1  &  {\cal A}_2  &  {\cal B}_1 &   {\cal S}_2  &  {\cal A}_1& 
  {\cal B}_2 &   {\cal C}_2 &   {\cal C}_1 \\\hline
 {\cal S}_1 & 0.62 & -0.01 & 0.07 & 0.25 &  &  &  & -0.02 \\
{\cal A}_2& -0.01 &  & -0.06 &  & 0.05 & 0.04 & 0.02 & -0.16 \\
{\cal B}_1 & 0.07 & -0.06 & -0.01 &  & 0.01 & 0.13 & -0.01 &  \\\hline
{\cal S}_2& 0.25 &  &  & 0.06 &  & &  &  \\
{\cal A}_1&  & 0.05 & 0.01 &  &  & -0.07 & -0.07 & 0.02 \\
{\cal B}_2 &  & 0.04 & 0.13 &  & -0.07 & -0.10 & -0.02 & 0.13 \\\hline
{\cal C}_2 &  & 0.02 & -0.01 &  & -0.07 & -0.02 & -0.11 & 0.37 \\\hline
{\cal C}_1 & -0.02 & -0.16 &  &  & 0.02 & 0.13 & 0.37 & -0.15
\end{array}\label{LSresult}
\end{equation} 
To illustrate how to read Eq.\,(\ref{LSresult}) we note that the largest single entry is associated with ${\cal S}_1 \otimes {\cal S}_1$, which represents the quark outside the scalar diquark correlation carrying all the nucleon's spin.  That is the $u$-quark in the proton.  However, it is noteworthy that a contribution of similar magnitude is associated with the axial-vector diquark correlations, expressing mixing between $p$- and $d$-waves; viz., ${\cal C}_1 \otimes {\cal C}_2 + {\cal C}_2 \otimes {\cal C}_1$.  With ${\cal C}_2$ all quark spins are aligned with that of the nucleon and the unit of angular momentum is opposed, while with ${\cal C}_1$ all quark spins are opposed and the two units of angular momentum are aligned.  This contribution is more important than those associated with ${\cal S}_2$; namely, scalar diquark terms with the bystander quark's spin antiparallel.  Finally, for the present, in this context one single number is perhaps most telling: the contribution to the normalization from $(L=0) \otimes (L=0)$ terms is only 37\% of the total.

\section{Coda}
The DSEs provide a natural vehicle for the exploration of confinement and DCSB.  DCSB is a remarkably effective mass generating mechanism.  For light-quarks it is far more important than the Higgs mechanism.  It is understood via QCD's gap equation, which delivers a quark mass function with a momentum-dependence that connects the perturbative domain with the nonperturbative, con\-sti\-tuent-quark domain.  The existence of a sensible truncation scheme enables the proof of exact results using the DSEs.  The scheme is also tractable.  Hence the results can be illustrated and predictions made for observables.  The consequent opportunities for rapid feedback between experiment and theory brings within reach an intuitive understanding of nonperturbative strong interaction phenomena.  

It can be argued that confinement is expressed in the analyticity properties of dressed Schwinger functions \cite{Roberts:1994dr}.  To build understanding it is essential to work toward an accurate map of the confinement force between light-quarks.  Among the rewards are a clear connection between confinement and DCSB, an accounting of the distribution of mass within hadrons, and a realistic picture of hybrids and exotics.

It is important to understand the relationship between parton properties on the light-front and the rest frame structure of hadrons.  This is a challenge because, e.g., DCSB, a keystone of low-energy QCD, has not been realized in the light-front formulation.  Parton distribution functions must be calculated in order to learn their content.  Parametrization is insufficient.  It would be very interesting to know how, if at all, the distribution functions of a Goldstone mode differ from those of other hadrons.  Answers to these and kindred questions are being sought using the DSEs \cite{Hecht:2000xa,Cloet:2007em}.

\subsection*{Acknowledgments}
CDR is grateful to the organizers for their hospitality and support.
This work was supported by the Department of Energy, Office of Nuclear Physics, contract no.\ DE-AC02-06CH11357; the Austrian Science Fund FWF under Schr\"odinger-R\"uckkehrstipendium Nr.~R50-N08; and benefited from the facilities of the ANL Computing Resource Center.


\begin{thebibliography}{00}
\bibitem{pmprivate} In preparing this table, we benefited from discussions with P.~Maris.
\end{thebibliography}
\end{document}